\documentclass[12pt,leqno]{article}
\title{Gravitational instability of polytropic spheres and generalized thermodynamics}
\author{Pierre-Henri Chavanis$^{1,2}$}
\date{}

\usepackage{amsthm,amsmath,amssymb,a4wide,psfig}
\def\mb#1{\setbox0=\hbox{$#1$}\kern-.025em\copy0\kern-\wd0
\kern-0.05em\copy0\kern-\wd0\kern-.025em\raise.0233em\box0}

\begin{document}
\maketitle
\vspace*{-1cm}
\begin{center}
$^{1}$ Laboratoire de Physique Quantique,
Universit\'e Paul Sabatier,\\
118 route de Narbonne 31062 Toulouse, France.\\

$^{2}$ Institute for Theoretical Physics,
University of California, Santa Barbara, California.\\

\vspace{0.5cm}
\end{center}

\begin{abstract}

We complete the existing literature on the structure and stability of
polytropic gas spheres reported in the classical monograph of
Chandrasekhar (1942). For isolated polytropes with index $1<n<5$, we
provide a new, alternative, proof that the onset of instability occurs
for $n=3$ and we express the perturbation profiles of density and
velocity at the point of marginal stability in terms of the Milne
variables. Then, we consider the case of polytropes confined within a
box of radius $R$ (an extension of the Antonov problem for isothermal
gas spheres). For $n\ge 3$, the mass-density relation presents some
damped oscillations and there exists a limiting mass above which no
hydrostatic equilibrium is possible. Like for isothermal gas spheres,
the onset of instability occurs precisely at the point of maximum mass
in the series of equilibrium. Analytical results are obtained for the
particular index $n=5$. We also discuss the relation of our study with
generalized thermodynamics (Tsallis entropy) recently investigated by
Taruya \& Sakagami (cond-mat/0107494).

\end{abstract}

\section{Introduction}
\label{sec_introduction}

In earlier papers (Chavanis 2001a,b) we investigated the gravitational
instability of finite isothermal spheres in Newtonian gravity and
general relativity. On a theoretical point of view, these systems
exhibit an interesting behavior with the occurence of phase
transitions associated with gravitational collapse. A rich stability
analysis follows and can be conducted analytically or by using graphical
constructions. On an astrophysical point of view, these studies can be
relevant for various systems including elliptical galaxies, globular
clusters, the interstellar medium or the core of neutron stars. In
this paper, we propose to extend our study to the case of polytropic
gas spheres. Polytropes with index $1<n<5$ are self-confined, so it is
not necessary to introduce an artifical ``box'' to limitate their
spatial extent. Using the methods developed for isothermal
configurations, we show that the transition from stability to
instability corresponds to an index $n=3$. This result is well-known
but we provide a new derivation based on the exact resolution of the
pulsation equation for polytropes. The perturbation profiles at the
point of marginal stability are expressed in terms of the Milne
variables. The density profile has only one node and the velocity
perturbation is proportional to the radial distance. Then, we consider
the case of polytropes with arbitrary index $n>1$ confined within a
box of radius $R$. For $n= \infty$, we recover the classical Antonov
(1962) problem for isothermal gas spheres.  However, for $n\ge 3$ we
already obtain results strikingly similar to those obtained for
isothermal configurations. In particular, confined polytropes in
hydrostatic equilibrium can exist only below a limiting mass (for a
given box radius $R$) and the series of equilibrium becomes unstable
precisely at the point of maximum mass. Subsequent oscillations in the
mass-density profile (for $n>5$) are associated with secondary modes
of instability. The locii of these modes of instability follow a
geometric progression with a ratio depending on the index of the
polytrope. For $n=\infty$, we recover the ratio $10.74...$ of
isothermal gas spheres. While this paper was in preparation, we came
accross the preprint of Taruya \& Sakagami (2001) on a related
subject. These authors investigate the stability of polytropes in the
framework of extended thermodynamics, using Tsallis
entropy. Therefore, in section \ref{sec_tsallis}, we analyze the
connexion of the present study with their approach and we discuss the
relevance of Tsallis entropy for describing astrophysical systems.

\section{Properties of polytropic gas spheres}
\label{sec_prop}

\subsection{The Lane-Emden equation}
\label{sec_eos}

Polytropic stars are characterized by an equation of state of the form 
\begin{equation}
p=K\rho^{\gamma},
\label{EOS1}
\end{equation} 
where $K$ and $\gamma$ are constants. The index $n$ of the polytrope is defined by the relation
\begin{equation}
\gamma=1+{1\over n}.
\label{EOS2}
\end{equation} 

The equation of state (\ref{EOS1}) corresponds to an adiabatic equilibrium in regions where convection keeps the star stirred up and produces a uniform entropy distribution ($s={\rm Cst.}$). In that case, $\gamma$ is the ratio of specific heats $c_{p}/c_{V}$ at constant pressure and volume. For a monoatomic gas, $\gamma=5/3$. The equation of state (\ref{EOS1}) also describes a polytropic equilibrium characterized by a uniform specific heat $c\equiv \delta Q/dT$. In this more general situation $\gamma=(c_{p}-c)/(c_{V}-c)$. Convective equilibrium is recovered for $c=0$ and isothermal equilibrium is obtained in the limit of infinite specific heat $c\rightarrow +\infty$. A power law relation between pressure and density is also the limiting form of the equation of state describing a gas of cold degenerate fermions (Chandrasekhar 1942). In that case, the constant $K$ can be expressed in terms of fundamental constants. In the classical limit $\gamma=5/3$, $n=3/2$ and $K={1\over 20}(3/\pi)^{2/3}h^{2}/ m$ (where $h$ is the Planck constant) and in the relativistic limit $\gamma=4/3$, $n=3$ and $K={1\over  8}(3/\pi)^{1/3}hc$ (where $c$ is the speed of light). Historically, the index $5/3$ appears in the classical theory of white dwarf stars initiated by Fowler (1926) and the index $4/3$ is related to the limiting mass of Chandrasekhar (1931). 

The condition of hydrostatic equilibrium of a spherically distribution of matter reads
\begin{equation}
{dp\over dr}=-\rho{d\Phi\over dr},
\label{EOS3}
\end{equation} 
where $\Phi$ is the gravitational potential. Using the Gauss theorem
\begin{equation}
{d\Phi\over dr}={GM(r)\over r^{2}},
\label{EOS4}
\end{equation} 
where $M(r)=\int_{0}^{r}\rho 4\pi r^{2}dr$ is the mass contained within the sphere of radius $r$, we can derive the fundamental equation of equilibrium (Chandrasekhar 1942)
\begin{equation}
{1\over r^{2}}{d\over dr}\biggl ({r^{2}\over\rho}{dp\over dr}\biggr )=-4\pi G\rho.
\label{EOS5}
\end{equation}  
Eqs. (\ref{EOS1})(\ref{EOS5}) fully determine the structure of polytropic gas spheres. Letting
\begin{equation}
\rho=\rho_{0}\theta^{n},\qquad \xi=\biggl\lbrack {4\pi G\rho_{0}^{1-1/n}\over K(n+1)}\biggr\rbrack^{1/2}r,
\label{EOS6}
\end{equation} 
where $\rho_{0}$ is the central density, we can reduce the condition of hydrostatic equilibrium to the Lane-Emden equation (Chandrasekhar 1942)
\begin{equation}
{1\over\xi^{2}}{d\over d\xi}\biggl (\xi^{2}{d\theta\over d\xi}\biggr )=-\theta^{n},
\label{EOS7}
\end{equation} 
with boundary conditions 
\begin{equation}
\theta(0)=1,\qquad \theta'(0)=0. 
\label{EOS8}
\end{equation} 

\begin{figure}[htbp]
\centerline{
\psfig{figure=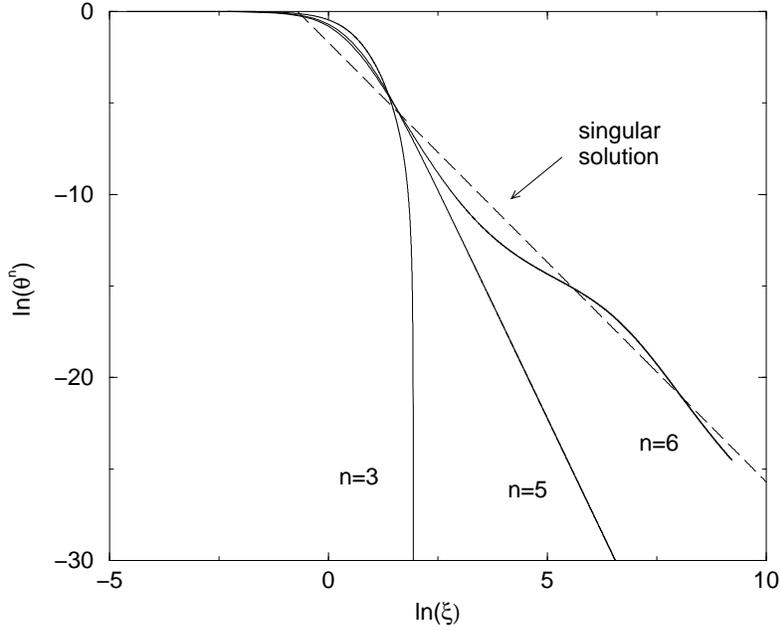,angle=0,height=8.5cm}}
\caption{Density profiles of polytropes  with index $n=3,5,6$. The dash line corresponds to the singular solution (\ref{sing}).   }
\label{profiles}
\end{figure}

For $n>3$, the Lane-Emden equation admits an analytical solution which is singular at the origin:
\begin{equation}
\theta_{s}=\biggl\lbrack {2(n-3)\over (n-1)^{2}}\biggr \rbrack^{1\over n-1}{1\over \xi^{2\over n-1}}.
\label{sing}
\end{equation}
Regular solutions of the Lane-Emden equation must in general be computed numerically. For $\xi\rightarrow 0$, we can expand the function $\theta$ in Taylor series. The first terms in this expansion are given by
\begin{equation}
\theta=1-{1\over 6}\xi^{2}+{n\over 120}\xi^{4}+...
\label{EOS9}
\end{equation} 
The behavior of $\theta(\xi)$ at large distances deserves a more specific discussion. For $1<n<5$, the density falls off to zero at a finite radius $R$, identified as the radius of the star. We denote by $\xi_{1}$ the value of the normalized distance at which $\theta=0$. For $\xi\rightarrow \xi_{1}$, we have
\begin{equation}
\theta=-\xi_{1}\theta'_{1}\biggl \lbrack {\xi_{1}-\xi\over\xi_{1}}+\biggl ({\xi_{1}-\xi\over\xi_{1}}\biggr )^{2}+\biggl ({\xi_{1}-\xi\over\xi_{1}}\biggr )^{3}+...\biggr\rbrack.
\label{EOS10}
\end{equation} 
For $n>5$, the polytropes extend to infinity, like the isothermal configurations recovered in the limit $n=\infty$. For $\xi\rightarrow +\infty$,
\begin{equation}
\theta=\biggl\lbrack {2(n-3)\over (n-1)^{2}}\biggr \rbrack^{1\over n-1}{1\over \xi^{2\over n-1}}\biggl\lbrace 1+{C\over \xi^{n-5\over 2(n-1)}}\cos\biggl\lbrack {\sqrt{7n^{2}-22n-1}\over 2(n-1)}\ln\xi+\delta\biggr\rbrack\biggr\rbrace.
\label{EOS11}
\end{equation} 
The curve (\ref{EOS11}) intersects the singular solution (\ref{sing})
infinitely often at points that asymptotically increase geometrically
in the ratio $1:{\rm exp}\lbrace 2(n-1)\pi/\sqrt{7n^{2}-22n-1}\rbrace$.
Since $\theta^{n}\sim \xi^{-2n\over n-1}$ at large distances, these
configurations have an ``infinite mass'', which is clearly
unphysical. In the following, we shall restrict these configurations to a
``box'' of radius $R$, like in the classical Antonov
problem. Therefore, Eq. (\ref{EOS7}) must be solved for $\xi\le
\alpha$ with
\begin{equation}
\alpha=\biggl\lbrack {4\pi G\rho_{0}^{1-1/n}\over K(n+1)}\biggr\rbrack^{1/2}R.
\label{EOS12}
\end{equation} 
Note that, for a fixed box radius $R$, $\alpha$ is a measure of the central density $\rho_{0}$. The case $n=5$ is special. For this index, the Lane-Emden equation can be solved analytically and yields the result:
\begin{equation}
\theta={1\over (1+{1\over 3}\xi^{2})^{1/2}}.
\label{EOS13}
\end{equation} 
The total mass of this configuration is finite but its potential energy diverges. Therefore, this polytrope must also be confined within a box. On Fig. \ref{profiles}, we have ploted different density profiles corresponding to the polytropic indices $n=3,5,6$.

\subsection{The Milne variables}
\label{sec_milne}

\begin{figure}[htbp]
\centerline{
\psfig{figure=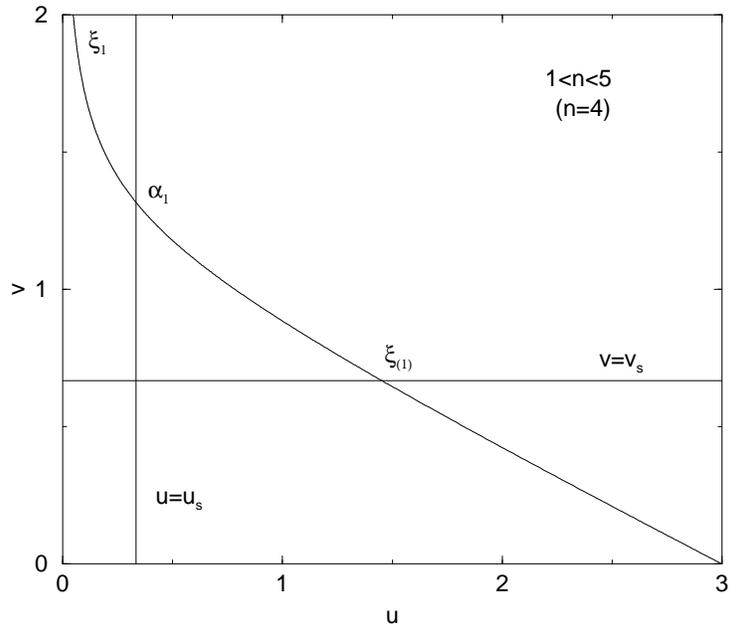,angle=0,height=8.5cm}}
\caption{The $(u,v)$ plane for polytropes with index $1<n<5$. The construction is made explicitly for $n=4$.}
\label{uv4}
\end{figure}

\begin{figure}[htbp]
\centerline{
\psfig{figure=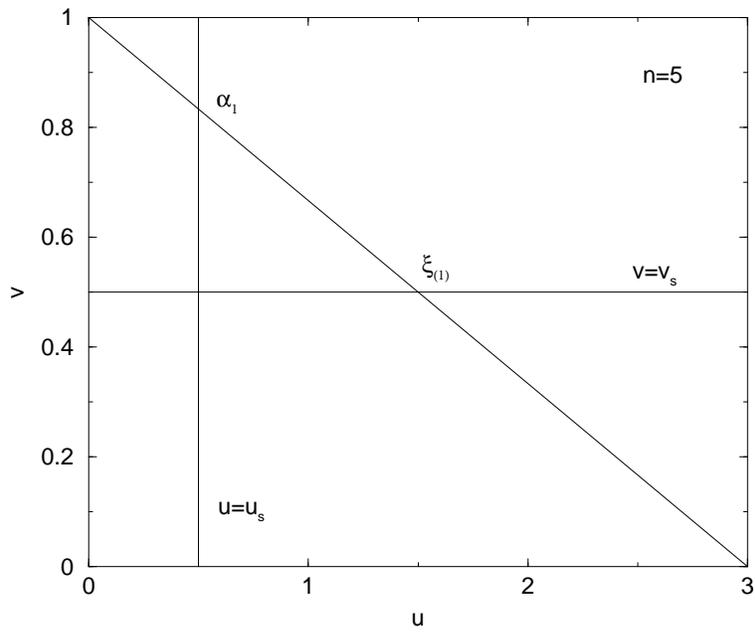,angle=0,height=8.5cm}}
\caption{The $(u,v)$ plane for polytropes with index $n=5$.}
\label{uv5}
\end{figure}

\begin{figure}[htbp]
\centerline{
\psfig{figure=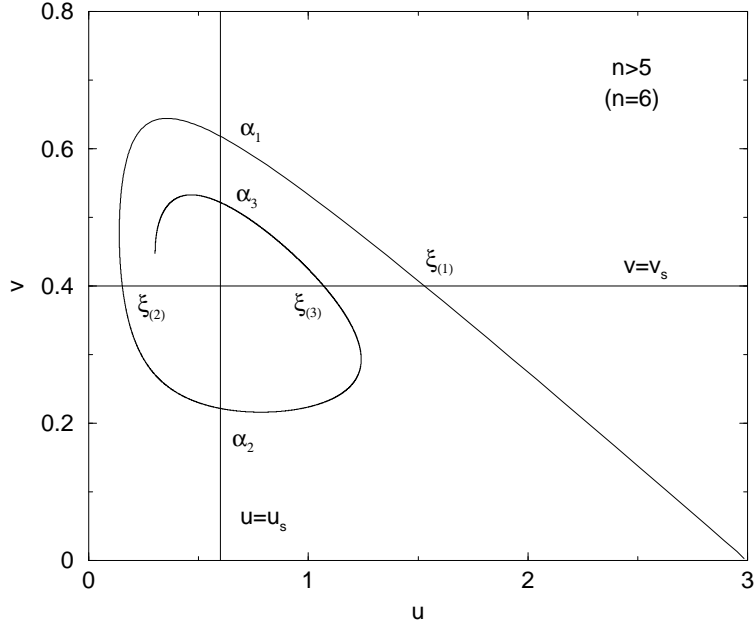,angle=0,height=8.5cm}}
\caption{The $(u,v)$ plane for polytropes with index $n>5$. The construction is made explicitly for $n=6$.}
\label{uv6}
\end{figure}

Like in the analysis of isothermal gas spheres, it will be convenient in the following to introduce the Milne variables $(u,v)$ defined by (Chandrasekhar 1942):
\begin{equation}
u=-{\xi\theta^{n}\over\theta'},\qquad v=-{\xi\theta'\over\theta}.
\label{uv}
\end{equation}
Taking the logarithmic derivative of $u$ and $v$ with respect to $\xi$ and using Eq. (\ref{EOS7}), we get
\begin{equation}
{1\over u}{du\over d\xi}={1\over \xi}(3-nv-u), 
\label{uv1}
\end{equation}
\begin{equation}
{1\over v}{dv\over d\xi}={1\over\xi}(u+v-1). 
\label{uv2}
\end{equation}
Due to the homology invariance of the polytropic configuations (see Chandrasekhar 1942), the Milne variables $u$ and $v$ satisfy a {\it first order} differential equation 
\begin{equation}
{u\over v}{dv\over du}=-{u+v-1\over u+nv-3}.
\label{uv3}
\end{equation}
The solution curve in the $(u,v)$ plane (see Figs. \ref{uv4}-\ref{uv6}) is parametrized by $\xi$. It starts from the point $(u,v)=(3,0)$ with a slope $(dv/du)_{0}=-{5\over 3n}$ as $\xi\rightarrow 0$. For $1<n<5$, the curve is monotonous and tends to $(u,v)=(0,+\infty)$ as $\xi\rightarrow \xi_{1}$. More precisely, using Eq. (\ref{EOS10}), we have 
\begin{equation}
uv^{n}\sim \omega_{n}^{n-1},\qquad \omega_{n}=-\xi_{1}^{n+1\over n-1}\theta'_{1} \qquad (\xi\rightarrow \xi_{1}).
\label{uv4a}
\end{equation}
For $n>5$, the solution curve spirals indefinitely around the fixed point $(u_{s},v_{s})=({n-3\over n-1}, {2\over n-1})$, corresponding to the singular solution (\ref{sing}), as $\xi$ tends to infinity. All polytropic spheres must necessarily lie on this curve. For bounded polytropic spheres, $\xi$ must be terminated at the box radius $\alpha$. For $n=5$, we have the following analytical relation between the Milne variables:
\begin{equation}
u+3v=3.
\label{uv5b}
\end{equation}

\subsection{The maximum mass and minimum temperature of confined polytropes}
\label{sec_limiting}

For polytropes confined within a box of radius $R$, there exists a well-defined relation between the mass $M$ of the configuration and the central density $\rho_{0}$ (through the parameter $\alpha$). Starting from the relation
\begin{equation}
M=\int_{0}^{R}\rho 4\pi r^{2}dr=4\pi\rho_{0}\biggl\lbrack {K(1+n)\rho_{0}^{1/n-1}\over 4\pi G}\biggr\rbrack^{3/2}\int_{0}^{\alpha}\theta^{n}\xi^{2}d\xi,
\label{L1}
\end{equation}  
and using the Lane-Emden equation (\ref{EOS7}), we get
\begin{equation}
M=-4\pi \rho_{0}\biggl\lbrack {K(1+n)\rho_{0}^{1/n-1}\over 4\pi G}\biggr\rbrack^{3/2}\alpha^{2}\theta'(\alpha).
\label{L2}
\end{equation}  
Expressing the central density in terms of $\alpha$, using Eq. (\ref{EOS12}), we obtain after some rearrangements
\begin{equation}
M=-4\pi\biggl\lbrack {K(1+n)\over 4\pi G}\biggr\rbrack^{n\over n-1} R^{n-3\over n-1}\alpha^{n+1\over n-1}\theta'(\alpha).
\label{L3}
\end{equation} 
Introducing the parameter 
\begin{equation}
\eta\equiv {M\over 4\pi}\biggl\lbrack { 4\pi G\over K(1+n)}\biggr\rbrack^{n\over n-1}{1\over R^{n-3\over n-1} }, 
\label{L4}
\end{equation} 
the foregoing relation can be rewritten
\begin{equation}
\eta=-\alpha^{n+1\over n-1}\theta'(\alpha).
\label{L5}
\end{equation}
For $n<5$, the normalized box radius $\alpha$ in necessarily
restricted by the inequality $\alpha\le\xi_{1}$. For the limiting
value $\alpha=\xi_{1}$, corresponding to an isolated polytrope satisfying
$\rho(R)=0$, we have
\begin{equation}
\eta(\xi_{1})=\omega_{n}.
\label{L5bis}
\end{equation}
The quantity $\omega_{n}$, defined by Eq. (\ref{uv4a}), is tabulated in
Chandrasekhar (1942).  The definition (\ref{L4}) of $\eta$ and the
relation (\ref{L5}) between $\eta$ and $\alpha$ are consistent with
the formulae derived in the case of an isothermal gas (see
Chavanis 2001a). This connexion is particularly relevant if we interpret
the constant $K$ as a polytropic temperature $\Theta_{\gamma}$ (see
Chandrasekhar 1942, p. 86). For $n\rightarrow +\infty$, we find that
$\eta\sim \eta_{\infty}/n$ where $\eta_{\infty}={\beta GMm\over R}$ is the 
normalized temperature of an isothermal sphere ($\beta=1/kT$).

\begin{figure}[htbp]
\centerline{
\psfig{figure=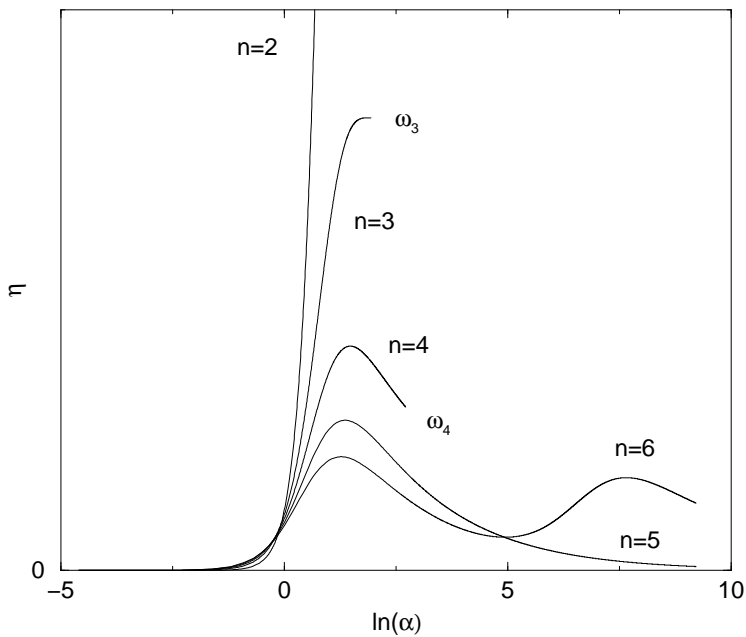,angle=0,height=8.5cm}}
\caption{Mass-density profiles for polytropic configurations with index $n=2,3,4,5,6$. A mass peak appears for the first time for the critical index $n=3$.  }
\label{alphaeta}
\end{figure}

The function $\eta(\alpha)$ is represented on Fig. \ref{alphaeta} for
different values of the polytropic index $n$. Instead of $\alpha$, we could have used the density contrast 
\begin{equation}
{\cal R}\equiv {\rho_{0}\over \rho(R)}=\theta(\alpha)^{-n}.
\label{contr}
\end{equation}  
Using the Lane-Emden equation (\ref{EOS7}) and the definition of the
Milne variables (\ref{uv}), it is straightforward to check that the
condition of extremum $d\eta/d\alpha=0$ is equivalent to
\begin{equation}
u_{0}={n-3\over n-1}\equiv u_{s},
\label{L6}
\end{equation}  
where, by definition, $u_{0}=u(\alpha)$ and $u_{s}$ refers to the singular solution (\ref{sing}). For $n\rightarrow \infty$, we recover the condition $u_{0}=1$ previously derived for isothermal configurations (Chavanis 2001a). The values of $\alpha$ for which $\eta$ is extremum are given by the intersection(s) between the solution curve in the $(u,v)$ plane and the line $u=u_{s}$ (see Figs. \ref{uv4}-\ref{uv6}). For $1<n<3$, $u_{s}<0$ and there is no intersection. The mass-density relation is therefore monotonous.  For $3\le n\le 5$, the curve $\eta(\alpha)$ presents a single maximum at 
$\alpha_{1}$. For $n=3$, this maximum is reached at the extremity of the curve ($\alpha_{1}=\xi_{1}$). For $n=5$, $\xi_{1}\rightarrow +\infty$ and the function $\eta(\alpha)$ is explicitely given by
\begin{equation}
\eta={\alpha^{5/2}\over 3(1+{1\over 3}\alpha^{2})^{3/2}}.
\label{L7}
\end{equation}  
The maximum of $\eta$ is located at $\alpha_{1}=\sqrt{15}$. Finally, for $n>5$, the mass-density relation presents an infinite number of damped oscillations since the line defined by Eq. (\ref{L6}) passes through the center of the spiral in the $(u,v)$ plane. If we denote by $\alpha_{k}$ the loci of the extrema of $\eta(\alpha)$, these values asymptotically follow the geometric progression 
\begin{equation}
\alpha_{k}\sim \biggl\lbrack e^{2(n-1)\pi\over \sqrt{7n^{2}-22n-1}}\biggr\rbrack^{k}\qquad (k\rightarrow \infty, \ {\rm integer}).
\label{L8}
\end{equation} 
obtained by substituting the asymptotic expansion (\ref{EOS11}) in Eq. (\ref{L6}). For $n\rightarrow\infty$, we recover the ratio $e^{2\pi/\sqrt{7}}=10.74...$ corresponding to classical isothermal configurations (Semelin {\it et al.} 1999, Chavanis 2001a).

From the above results, it is clear that  restricted polytropic spheres with index $n\ge 3$ can exist only for  
\begin{equation}
\eta\le -\alpha_{1}^{n+1\over n-1}\theta'(\alpha_{1})\equiv\eta_{max}.
\label{L9}
\end{equation}
This implies in particular the existence of a limiting mass (for a given confining  radius $R$) such that
\begin{equation}
M\le M_{max}=4\pi\eta_{max}\biggl\lbrack { K(1+n)\over 4\pi G}\biggr\rbrack^{n\over n-1} R^{n-3\over n-1}. 
\label{L10}
\end{equation}
Alternatively, for a given mass $M$ and radius $R$, this inequality implies the existence of a minimum value of the polytropic temperature $\Theta_{\gamma}=Km/k$. Indeed, $\Theta_{\gamma}$ is restricted by the inequality 
\begin{equation}
\Theta_{\gamma}\ge {4\pi Gm\over k(1+n)}\biggl ({M\over 4\pi \eta_{max}}\biggr )^{n-1\over n}{1\over R^{n-3\over n}}.
\label{L11}
\end{equation}
In the limit $n=\infty$, we recover the critical temperature $kT_{c}={GMm\over 2.52R}$ below which an isothermal sphere is expected to collapse (Lynden-Bell \& Wood 1968). This  minimum temperature, corresponding to $d\eta/d\alpha=0$,  appears for the first time for a critical index $n=3$. This observation will take a deeper physical significance in the stability analysis of the following section. For $n\le 3$, the mass-density profile is monotonous and  $\eta\le\omega_{n}$. The maximum mass (resp.  minimum temperature) of confined polytropes is always smaller (resp. larger) than the corresponding one for isolated polytropes. However, this bound does not correspond to a condition of extremum $d\eta/d\alpha=0$ but rather to the impossibility of constructing polytropes with $\alpha>\xi_{1}$ (since $\theta$ can become negative).

\section{Dynamical stability of polytropic gas spheres}
\label{sec_dyn}

\subsection{The equation of radial pulsations}
\label{sec_pulse}

We shall now investigate the dynamical stability of polytropic gas spheres against radial perturbations. We use a method and presentation similar to that adopted for isothermal spheres in Chavanis (2001a). The equations describing the motions of a gaseous star are the equation of continuity, the Euler equation and the Poisson equation  
\begin{equation}
{\partial \rho\over \partial t}+\nabla (\rho {\bf v})=0,
\label{F1}
\end{equation}
\begin{eqnarray}
{\partial {\bf v}\over \partial t}+({\bf v}\nabla){\bf v}=-{1\over\rho}\nabla p-\nabla\Phi,
\label{F2}
\end{eqnarray}
\begin{equation}
\Delta\Phi=4\pi G\rho.
\label{F3}
\end{equation}
It is possible to show that viscosity does not change the onset of instability. Therefore, for simplicity, we have directly used the Euler equation instead of the Navier-Stokes equation. We assume furthermore that the pressure and the density are connected by the polytropic equation of state (\ref{EOS1}). We write small perturbations around equilibrium in the form
\begin{equation}
{\bf v}=\delta {\bf v}({\bf r},t),\qquad \rho=\overline{\rho}+\delta \rho({\bf r},t),
\label{F5}
\end{equation}
\begin{equation}
p=\overline{p}+\delta p({\bf r},t),\qquad \Phi=\overline{\Phi}+\delta \Phi({\bf r},t),
\label{F6}
\end{equation}
where the bar refers to the stationary solution (in the following we shall drop the bar). The linearized equations for the perturbations are  
\begin{equation}
\rho{\partial {\delta{\bf v}}\over \partial t}=-K\gamma \nabla (\rho^{\gamma-1}\delta\rho)-\delta\rho\nabla\Phi-\rho\nabla\delta\Phi,
\label{d6}
\end{equation}
\begin{equation}
{\partial \delta \rho\over \partial t}+\nabla (\rho {\delta{\bf v}})=0,
\label{d7}
\end{equation}
\begin{equation}
\Delta\delta\Phi=4\pi G\delta\rho.
\label{d8}
\end{equation}
Restricting ourselves to radial perturbations and writing  the time dependance of the perturbation in the form $\delta v\sim e^{\lambda t}$, $\delta\rho\sim e^{\lambda t}$,..., the equations of the problem become
\begin{equation}
\lambda\rho\delta v=-K\gamma{d\over dr}(\rho^{\gamma-1}\delta\rho)-\delta\rho{d\Phi\over dr}-\rho{d\delta\Phi\over dr},
\label{d9}
\end{equation}
\begin{equation}
\lambda \delta \rho+ {1\over r^{2}}{d\over dr}(\rho r^{2}\delta{v})=0,
\label{d10}
\end{equation}
\begin{equation}
 {1\over r^{2}}{d\over dr}\biggl ( r^{2}{d\delta\Phi\over dr}\biggr )=4\pi G\delta\rho.
\label{d11}
\end{equation}
We introduce the function $q(r)$ by the relation
\begin{equation}
\delta\rho={1\over 4\pi r^{2}}{dq\over dr}.
\label{qr}
\end{equation}
Physically, $q(r)$ represents the mass perturbation $\delta M(r)=\int_{0}^{r}\delta\rho 4\pi r^{2}dr$ within the sphere of radius $r$. Thus, by definition, $q(0)=0$. Then, it is readily seen that the Poisson equation (\ref{d11}) is equivalent to
\begin{equation}
{d\delta\Phi\over dr}={Gq\over r^{2}},
\label{qrw}
\end{equation}
which is just the perturbed Gauss theorem (\ref{EOS4}). On the other hand, the continuity equation (\ref{d10}) leads to the relation 
\begin{equation}
\delta v=-{\lambda\over 4\pi \rho r^{2}}q.
\label{d13}
\end{equation}
Substituting these results back into Eq. (\ref{d9}), we obtain
\begin{equation}
 {\lambda^{2}\over 4\pi r^{2}}q=K\gamma{d\over dr}\biggl (\rho^{\gamma-1}{1\over 4\pi r^{2}}{dq\over dr}\biggr )+{1\over 4\pi r^{2}}{dq\over dr}{d\Phi\over dr}+{G\rho\over r^{2}}q.
\label{d14}
\end{equation}
Using the condition of hydrostatic equilibrium (\ref{EOS3}), the foregoing equation can be rewritten
\begin{equation}
K\gamma {d\over dr}\biggl ({\rho^{\gamma-2}\over 4\pi  r^{2}}{dq\over dr}\biggr )+{Gq\over  r^{2}}={\lambda^{2}\over 4\pi \rho r^{2}}q.
\label{d15}
\end{equation}
which  is the required equation of radial pulsations for a polytrope.

\subsection{The condition of marginal stability}
\label{sec_marginal}

Considering the case of marginal stability $\lambda=0$ and introducing the dimensionless variables defined in section \ref{sec_prop}, we can reduce the equation of radial pulsations to the form
\begin{equation}
{d\over d\xi}\biggl ({\theta^{1-n}\over\xi^{2}}{dF\over d\xi}\biggr )+{nF\over\xi^{2}}=0
\label{m1}
\end{equation}
with $F(0)=0$. Denoting by ${\cal L}$ the differential operator appearing in Eq. (\ref{m1}) and using the Lane-Emden equation (\ref{EOS7}), we easily establish that 
\begin{eqnarray}
{\cal L}(\xi^{2}\theta')={d\over d\xi}\biggl ({\theta^{1-n}\over\xi^{2}}{d\over d\xi}(\xi^{2}\theta')\biggr )+n\theta'
=-{d\over d\xi}(\theta^{1-n}\times \theta^{n})+n\theta'=(n-1)\theta',
\label{m2}
\end{eqnarray} 
\begin{eqnarray}
{\cal L}(\xi^{3}\theta^{n})={d\over d\xi}\biggl ({\theta^{1-n}\over\xi^{2}}{d\over d\xi}(\xi^{3}\theta^{n})\biggr )+n\xi\theta^{n}
=(3+n)\theta'+\xi n\theta''+n\xi\theta^{n}
=(3-n)\theta'.
\label{m3}
\end{eqnarray}
Therefore, the general solution of Eq. (\ref{m1}) is
\begin{equation}
F(\xi)=c_{1}\biggl (\xi^{3}\theta^{n}+{n-3\over n-1}\xi^{2}\theta'\biggr ),
\label{m4}
\end{equation} 
where $c_{1}$ is an arbitrary constant. The connexion with isothermal configurations (see Chavanis 2001a) is particularly obvious if we make the correspondance $\theta'\leftrightarrow \psi'$ and $\theta^{n}\leftrightarrow e^{-\psi}$.

\subsection{Boundary conditions}
\label{sec_bc}

The equation of pulsations (\ref{d15}) must be supplemented by appropriate boundary conditions. For self-confined polytropes ($1<n<5$), we require that the Lagrangian derivative of the pressure vanishes at the surface of the configuration, i.e.
\begin{equation}
{d\over dt}\delta p+\delta v{dp\over dr}=0, \qquad {\rm at}\  r=R.
\label{b1}
\end{equation} 
Using Eqs. (\ref{qr})(\ref{d13}), this condition can be rewritten 
\begin{equation}
{p\over\rho}\biggl ({dq\over dr}-{q\over\rho}{d\rho\over dr}\biggr )=0, \qquad {\rm at}\  r=R,
\label{b2}
\end{equation}
or, in dimensionless form, 
\begin{equation}
\theta{dF\over d\xi}-n{d\theta\over d\xi}F=0, \qquad {\rm at}\  \xi=\xi_{1}. 
\label{b3}
\end{equation} 
The index of the marginally stable polytrope is obtained by substituting the solution (\ref{m4}) in the boundary condition (\ref{b3}). Since $\theta(\xi_{1})=0$, this yields $n=3$. Therefore, the transition from stability to instability occurs for a polytropic index $n=3$ or, equivalently, for an adiabatic index $\gamma=4/3$. Of course, this result is well-known and could have been obtained from the general theorems of stellar pulsation (see, e.g., Cox 1980). However, we provide here an alternative method based on the explicit resolution of the equation of pulsations for polytropes.

\begin{figure}[htbp]
\centerline{
\psfig{figure=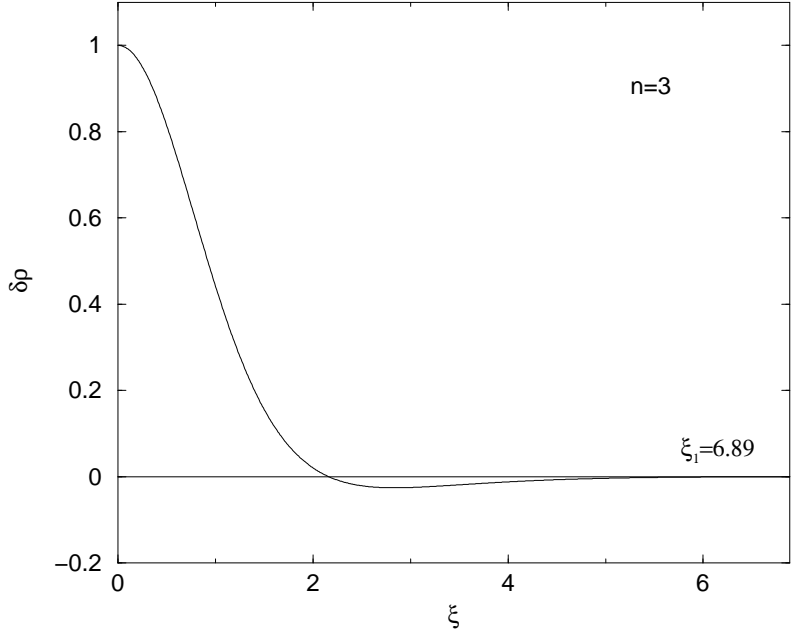,angle=0,height=8.5cm}}
\caption{Density perturbation profile for the critical  polytrope of index 
$n=3$.  }
\label{deltarho3}
\end{figure}

According to Eq. (\ref{qr}), the perturbation profile at the point of marginal stability is given by
\begin{equation}
{\delta\rho\over\rho_{0}}={1\over 4\pi\xi^{2}}{dF\over d\xi},
\label{b4}
\end{equation} 
with the expression (\ref{m4}) for $F(\xi)$. Simplifying the derivative with the aid of the Lane-Emden equation (\ref{EOS7}), we obtain
\begin{equation}
{\delta\rho\over\rho}={3\over 4\pi}c_{1}\biggl ({2\over n-1}-v\biggr ),
\label{b5}
\end{equation} 
where $v$ is the Milne variable (\ref{uv}). For $n=3$, it reduces to
\begin{equation}
\delta\rho={3\over 4\pi}c_{1}\rho_{0}\theta^{3}(1-v).
\label{b5bis}
\end{equation} 
This perturbation profile is ploted on Fig. \ref{deltarho3}.
On the other hand, the velocity profile is given by Eq. (\ref{d13}). Introducing the velocity of sound $c_{s}$, we find that
\begin{equation}
{\delta v\over c_{S}}=-{\lambda'\over 4\pi}c_{1}{\theta'\over\theta^{n}}\biggl ({n-3\over n-1}-u\biggr ).
\label{b6}
\end{equation} 
For $n=3$, we get
\begin{equation}
{\delta v\over c_{S}}=-{\lambda'\over 4\pi}c_{1}\xi,
\label{b7}
\end{equation} 
and we observe that the velocity is proportional to the radial distance. 

For polytropes confined within a box, the boundary condition consistent with the conservation of mass is $F(\alpha)=0$. With the expression  (\ref{m4}) for $F(\xi)$, this yields 
\begin{equation}
u_{0}={n-3\over n-1}
\label{b8}
\end{equation} 
which is precisely Eq. (\ref{L6}). Therefore, the point of marginal stability coincides with the point of maximum mass (or minimum temperature) in the series of equilibrium. For $n<3$, restricted polytropes are always stable since the function $\eta(\alpha)$ is monotonous. For $3\le n\le 5$, the configurations with $\alpha> \alpha_{1}$ are unstable but there are no secondary instabilities. For $n>5$, a first mode of stability is lost for $\alpha=\alpha_{1}$. Subsequent oscillations in the mass-density profile are associated with secondary  instabilities. 

The profiles of density and velocity that trigger these instabilities can be written
\begin{equation}
{\delta\rho\over\rho}={3\over 4\pi}c_{1}(v_{s}-v ),
\label{b9}
\end{equation} 
\begin{equation}
{\delta v\over c_{S}}=-{\lambda'\over 4\pi}c_{1}{\theta'\over\theta^{n}}(u_{s}-u).
\label{b10}
\end{equation} 
where we recall that $\xi\le\alpha_{k}$ for the mode of order $k$. We can determine the qualitative behavior of these profiles graphically by considering the intersections between the solution curve in the $(u,v)$ plane and the lines $v=v_{s}$ and $u=u_{s}$ (see Figs. \ref{uv4}-\ref{uv6}). These intersections determine the nodes of the perturbation profile. For $n\le 5$, there is only one intersection since the $(u,v)$ curve is monotonous. Therefore, the density profile $\delta\rho/\rho$ has only one node at $\xi_{(1)}=2.16...$ (see Fig. \ref{deltarho3}). For $n=5$, we obtain the analytical solutions 
\begin{equation}
{\delta\rho\over\rho}={3\over 8\pi}c_{1}{3-\xi^{2}\over 3+\xi^{2}},
\label{b11}
\end{equation} 
\begin{equation}
{\delta v\over c_{S}}={\lambda'\over 72\pi}c_{1}\xi(\xi^{2}-15).
\label{b12}
\end{equation}
The perturbation $\delta\rho$ becomes zero at $\xi_{(1)}=\sqrt{3}$.  
For $n>5$, there are several intersections with the spiral since the lines  $v=v_{s}$ and $u=u_{s}$ pass through the center of the spiral. The description of the perturbation profiles is  similar to the one given for isothermal spheres. In particular, the fundamental mode of instability has only one node and  high order modes of instability present numerous oscillations whose nodes follow a geometric progression. We refer to our previous paper (Chavanis 2001a) for a more precise description of these results.

\section{Generalized thermodynamics and Tsallis entropy}
\label{sec_tsallis}

Recently, Taruya \& Sakagami (2001) have investigated the stability of polytropic spheres within the framework of generalized thermodynamics. It has been argued by Tsallis and co-workers that ordinary statistical mechanics and thermodynamics does not describe correctly systems with long-range interactions, which are by essence non extensive. A family of functionals of the form 
\begin{equation}
S_{q}=-{1\over q-1}\int (f^q-f)d^{3}{\bf r}d^{3}{\bf v},
\label{t1}
\end{equation}  
known as Tsallis entropies, has been introduced to extend the classical Boltzmann-Gibbs statistical mechanics to these systems. These functionals are labeled by a parameter $q$.  The Boltzmann entropy is recovered in the limit $q\rightarrow 1$.  This new formalism has been applied in various domains of physics, astrophysics, fluid mechanics, biology, economy etc... and it extends the results obtained with ordinary statistical mechanics. In the case of self-gravitating systems, it was shown by Plastino \& Plastino (1993) that the extremalization of the Tsallis entropy at fixed mass and energy yields a polytropic equation of state of the form (\ref{EOS1}). The parameter $q$ is related to the index $n$ of the polytrope by the relation
\begin{equation}
n={1\over q-1}+{3\over 2}, \qquad (q\ge 1).
\label{t2}
\end{equation}
Then, Taruya \& Sakagami (2001) investigated the stability problem in
the microcanonical ensemble by extending the analysis of Padmanabhan
(1989) for the classical Antonov instability. Polytropic
configurations are said to be stable if they correspond to
(generalized) entropy {\it maxima} at fixed mass $M$ and energy $E$.
Taruya \& Sakagami showed that, for $n>5$, an equilibrium exists only
above a critical energy depending on the index of the
configuration. In the limit $n\rightarrow +\infty$, the Antonov result
$E_{c}=-0.335GM^{2}/R$ obtained with the Boltzmann entropy is
recovered. Furthermore, they showed that the onset of instability
coincides with the point of minimum energy in the series of
equilibrium in agreement with standard turning point analysis. For
$n<5$, there is no critical value of energy and the polytropic
configurations are stable. These results differ from those obtained in
the present paper where it is found that the index marking the
transition from stability to instability is $n=3$ (in agreement with
standard theorems of stellar pulsations).  The origin of this
discrepency is certainly related to the inequivalence of the
statistical ensembles in extended thermodynamics, like in ordinary
thermodynamics, for self-gravitating systems (see, e.g., Padmanabhan
1990). Taruya \& Sakagami work in the microcanonical ensemble and
maximize Tsallis entropy at fixed mass and energy. On the contrary, in
our study, we keep the polytropic temperature $\Theta_\gamma=Km/k$
constant, which corresponds to the canonical description. It would be
of interest to study the maximization of Tsallis free energy at fixed
mass and temperature to see if it provides the same conditions of
stability as our dynamical approach based on the Navier-Stokes
equations. For the Boltzmann entropy, a clear connexion was found
between dynamical stability and thermodynamical stability in the
canonical ensemble (Semelin {\it et al.} 2001, Chavanis 2001a) and it
is desirable to check whether this connexion is preserved by Tsallis
generalized thermodynamics. This requires in particular to define
properly the notion of temperature and free energy in the generalized
sense.

\begin{figure}[htbp]
\centerline{
\psfig{figure=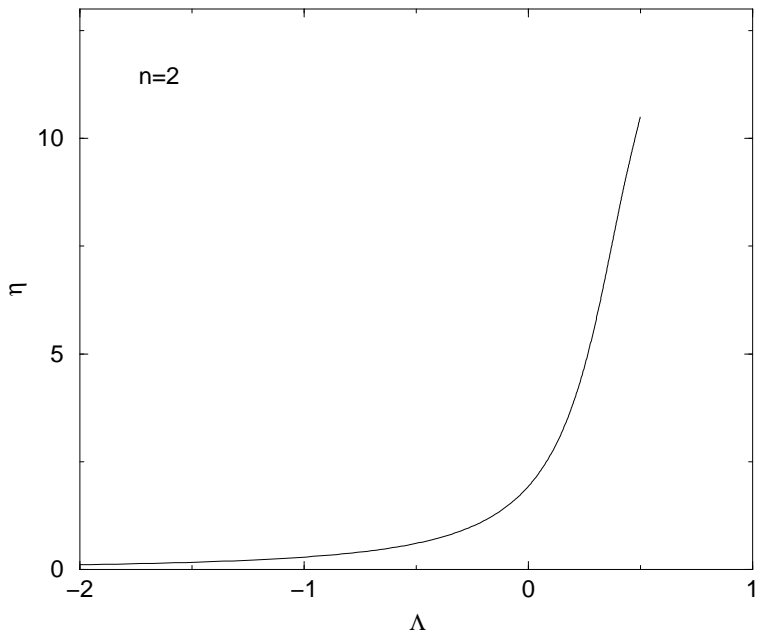,angle=0,height=8.5cm}}
\caption{Phase diagram for $n=2$.  }
\label{el2}
\end{figure}

\begin{figure}[htbp]
\centerline{
\psfig{figure=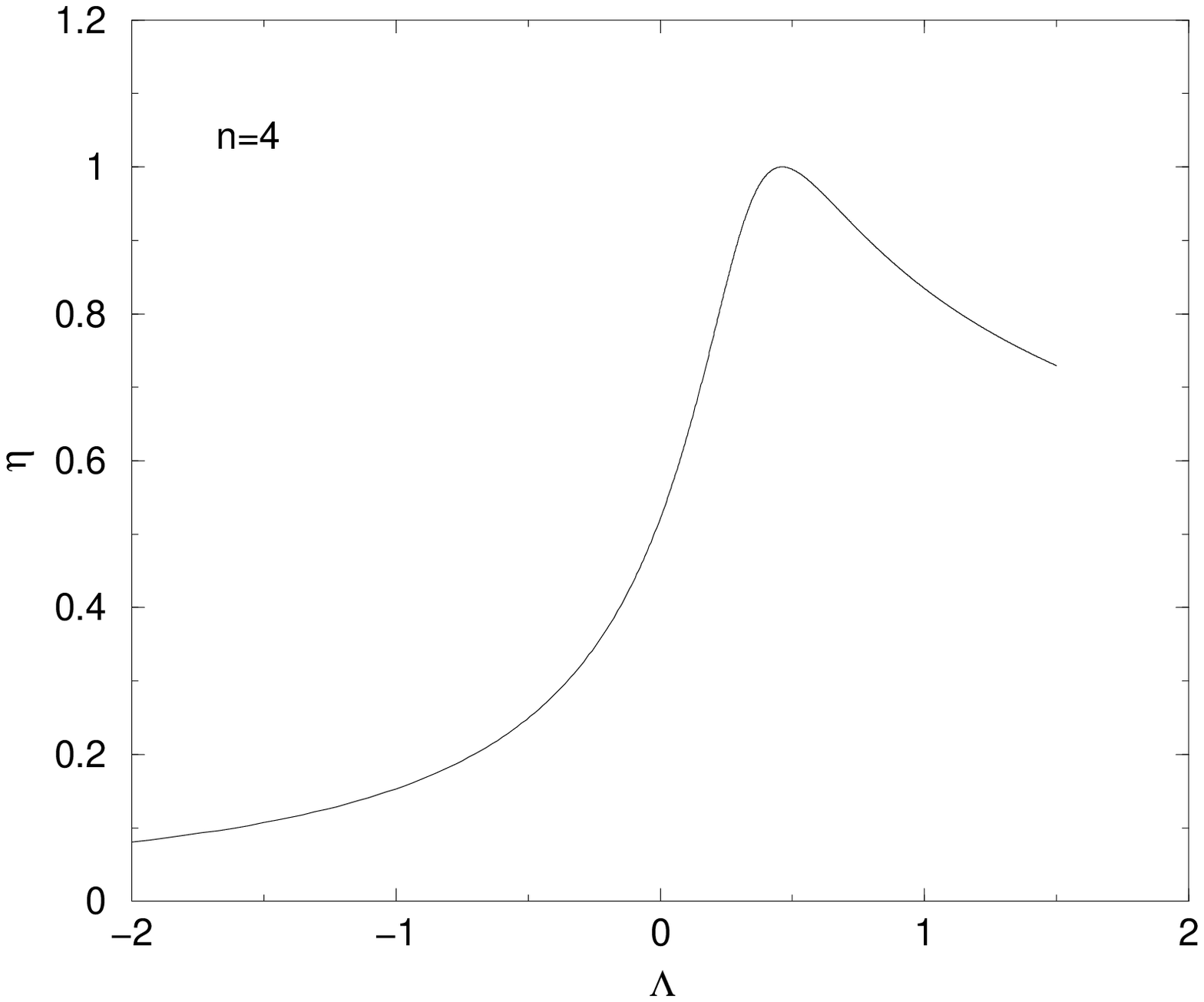,angle=0,height=8.5cm}}
\caption{Phase diagram for $n=4$.  }
\label{el4}
\end{figure}

\begin{figure}[htbp]
\centerline{
\psfig{figure=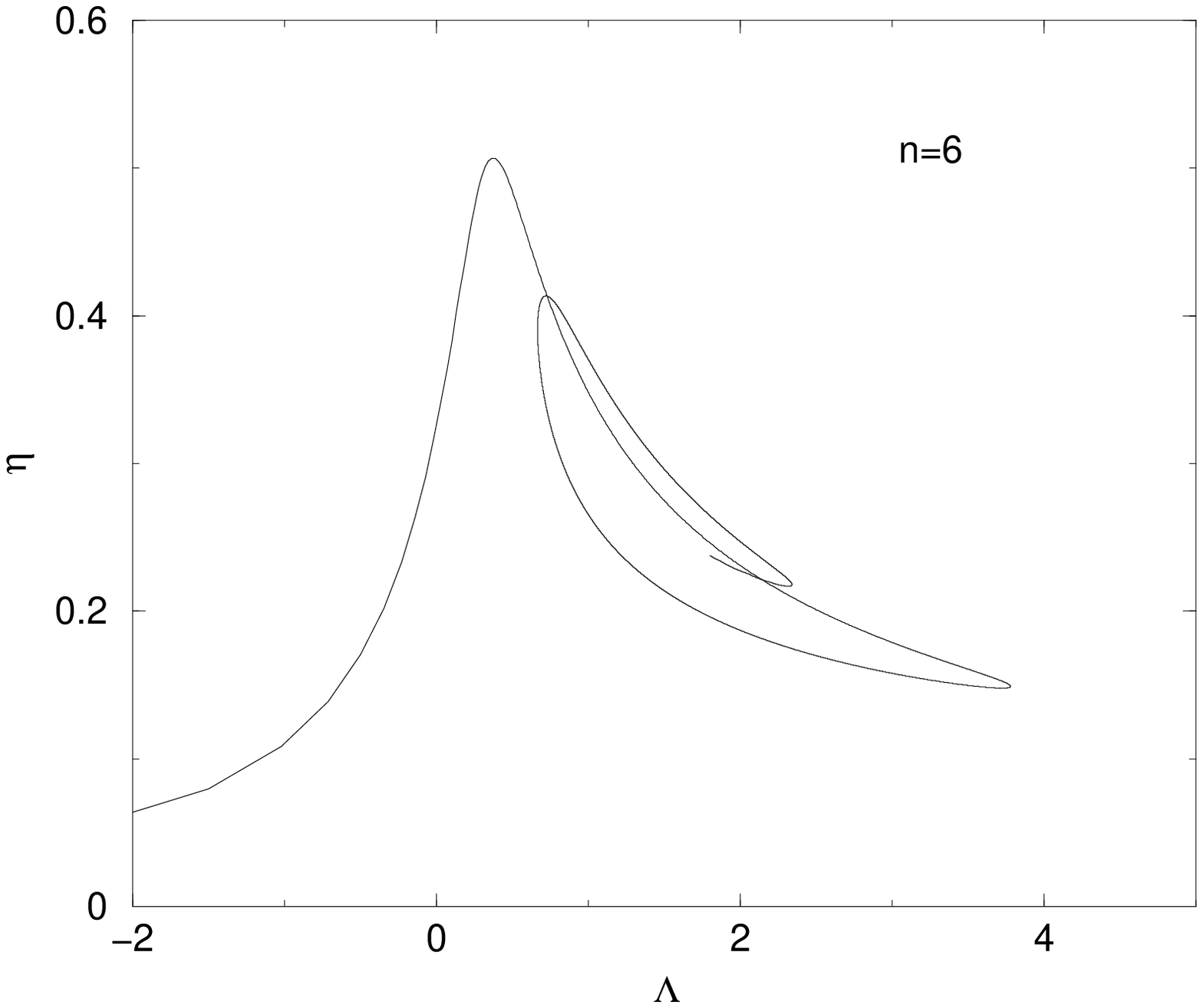,angle=0,height=8.5cm}}
\caption{Phase diagram for $n=6$.  }
\label{el6}
\end{figure}

\begin{figure}[htbp]
\centerline{
\psfig{figure=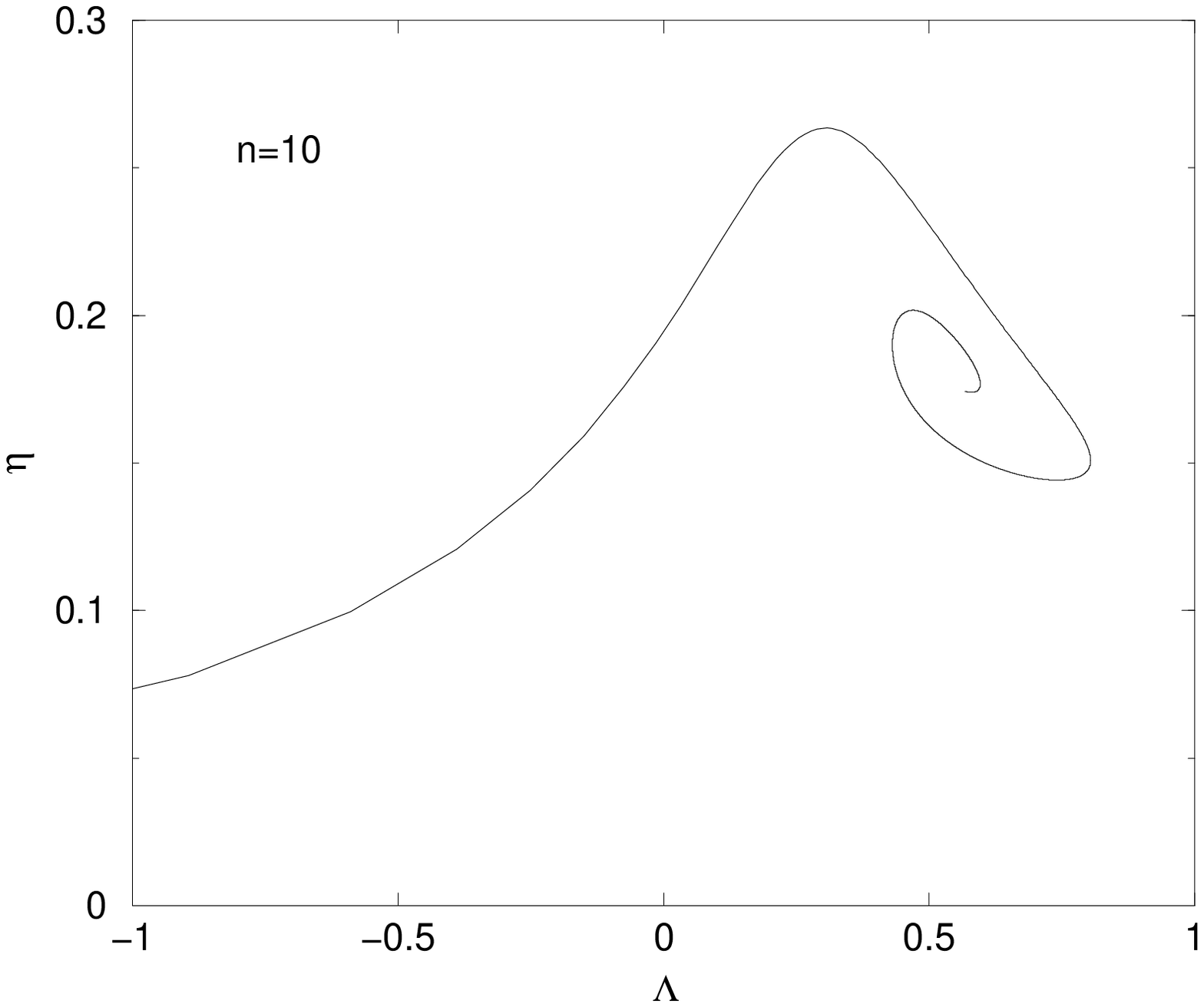,angle=0,height=8.5cm}}
\caption{Phase diagram for $n=10$.  }
\label{el10}
\end{figure}

On Figs. \ref{el2}-\ref{el10}, we have represented the equilibrium
phase diagram (for different values of $n$) obtained in the framework
of extended thermodynamics, interpreting the parameter $\eta$ as a generalized
inverse temperature. The temperature-energy curve is defined
in a parametric form by the equations
\begin{equation}
\eta=-\alpha^{n+1\over n-1}\theta'(\alpha)
\label{t3}
\end{equation} 
\begin{equation}
\Lambda\equiv -{ER\over GM^{2}}=-{1\over n-5}\biggl\lbrack {3\over 2}\biggl (1-{1\over v(\alpha)}\biggr )+{n-2\over n+1}{u(\alpha)\over v(\alpha)}\biggr\rbrack
\label{t4}
\end{equation}
The expression (\ref{t4}) for the energy has been derived by Taruya \& Sakagami (2001). For $n<3$, there is no turning point in the diagram, so the polytropes are always stable. For $3<n<5$, the inverse temperature $\eta$ presents a maximum but not the energy. Therefore, the polytropes are always stable in the microcanonical ensemble but they are unstable in the canonical ensemble after the turning point. It can be noted that the unstable region has a negative specific heat $C\sim {d\Lambda\over d\eta}<0$ (in the generalized sense). For $n>5$, the temperature and the energy both present an infinite number of extrema. For $n\sim 6$, there are crossing points at which two solutions have the same value of temperature and energy (but a different density contrast). The polytropes are unstable in the canonical ensemble after the first turning point of temperature and they are unstable in the microcanonical ensemble after the first turning point of energy. These results are strikingly similar to those obtained with the classical Boltzmann entropy $(n=\infty)$.

It is interesting that the results obtained for isothermal gas spheres
described by ordinary thermodynamics (Boltzmann entropy) can be
extended to polytropic configurations by introducing a generalized
functional (Tsallis entropy). This makes this new formalism of
interest. However, we would like to make two comments. (i) First of all,
Tsallis entropy does {\it not} describe polytropic gaseous stars
although it yields an equation of state of the polytropic form. Indeed,
Tsallis entropy predicts {\it in addition} a distribution function of
the form
\begin{equation}
f({\bf r},{\bf v})=A\biggl \lbrack \Phi_{0}-\Phi({\bf r})-{1\over 2}v^{2}\biggr\rbrack^{1/(q-1)},
\label{t5}
\end{equation} 
whereas in polytropic stars, the distribution function corresponds to a local thermodynamical equilibrium (L.T.E) characterized by a Maxwellian distribution of velocities
\begin{equation}
f({\bf r},{\bf v})=\biggl ({m\over 2\pi kT({\bf r})}\biggr )^{3/2}\rho({\bf r})e^{-{mv^{2}\over 2kT({\bf r})}},
\label{t6}
\end{equation} 
with a local temperature $T({\bf r})$ (recall that, for convective
equilibrium, the entropy is uniform not the temperature).  In view of
these remarks, the connexion between (generalized) thermodynamical
stability (in the canonical ensemble) and dynamical stability with
respect to Navier-Stokes equations is not obvious. It is however
expected to be true since the condition of dynamical stability is
consistent with the turning point argument of thermodynamical
stability. (ii) Tsallis entropy could describe a particular class of
stellar systems, apparently not observed in nature, known as stellar
polytropes (see Binney \& Tremaine 1987). These systems have the same
structure as polytropic gas spheres in physical space but not in phase
space (the equivalence only occurs for a polytropic index $n=\infty$,
i.e. for an isothermal configuration). This distinction is important
because isolated stellar polytropes are always stable with respect to
the Vlasov (or collisionless Boltzmann) equation (see Binney \&
Tremaine 1987) while isolated polytropic stars are unstable with
respect to the Navier-Stokes equations for $n\ge 3$. Generalized
thermodynamics could be developed (at least formally) for this particular
class of stellar systems.

However, generalized thermodynamics is often presented as a way of
solving the ``problems'' associated with the use of the Boltzmann
entropy in systems with long-range interactions and we would like to
criticize this interpretation. For self-gravitating systems, the
``problems'' usually reported are the absence of global entropy
maximum (resulting in the ``infinite mass problem'' and the
``gravothermal catastrophe''), the negative specific heats, the
inequivalence of statistical ensembles and the absence of
thermodynamical limit (a discussion of these concepts can be found in
Chavanis {\it et al.} 2001). It should be noted that the Tsallis
entropy displays the same phenomena, at least for values of $q<9/7$
(i.e. $n>5$) considered by Taruya \& Sakagami (2001). On the other
hand, these relatively unusual phenomena should not cause surprise if
one recognizes that self-gravitating systems are relatively
exceptional among $N$-body systems. For example the ``infinite mass
problem'' (see, e.g., Binney
\& Tremaine 1987) is just a mathematical curiosity. 
There is no justification in maximizing the entropy in an infinite
domain even if an entropy maximum exists. Due to kinetic effects, the
relaxation is necessarily {\it incomplete} and the ergodic hypothesis
which sustains the statistical analysis is  restricted to a
{\it finite} region of space. This incomplete relaxation is
qualitatively discussed by Lynden-Bell (1967) in his statistical
description of the ``violent relaxation'' of collisionless stellar
systems (e.g., elliptical galaxies). A similar limitation occurs in
the context of two-dimensional turbulence to understand the
confinement of vortices that form after a rapid merging (see, e.g.,
Chavanis \& Sommeria 1998 and in particular Brands {\it et al.}  1999
for a discussion of Tsallis entropy). This incomplete relaxation can
be described by kinetic equations of the form given by Chavanis {\it
et al.} (1996) with a space dependant diffusion coefficient. A
convincing numerical evidence of this kinetic confinement was given by
Robert \& Rosier (1997) in two-dimensional turbulence.  In addition,
in astrophysics, specific processes (e.g., the evaporation of stars,
tidal effects,...)  must be taken into account and can restrict the
domain of applicability of statistical mechanics. These limitations
are usually accounted for by introducing truncated models like the
Michie-King model for globular clusters or the models proposed by
Stiavelli \& Bertin (1987) or Hjorth \& Madsen (1993) for elliptical
galaxies. These models lead to composite configurations with an
isothermal core and a polytropic envelope.  With these modifications
(justified by precise physical arguments), ordinary statistical
mechanics provides in general a good explanation of astrophysical
phenomena and is consistent with observations. Note that the
above-mentioned truncated models differ from pure polytropes, so that
they cannot be justified by generalized thermodynamics
\footnote{Recently, Fa
\& Pedron (2001) have proposed to describe elliptical galaxies by an
extended King model using Tsallis entropy. They obtained a good fit
for NGC 3379.  However, we beleive that this agreement is
fortuitous. An even better fit has been obtained by Hjorth \& Madsen
(1993) with a different model justified by more convincing physical
arguments. Their model has an isothermal (instead of a polytropic)
inner region in agreement with Lynden-Bell's statistical mechanics of
violent relaxation. The polytropic envelope is justified by incomplete
relaxation (or equivalently by a lack of ergodicity) when we depart
from the central mixing region. Their model differs from the
Michie-King model because the system is collisionless and not subject
to tidal forces (King's models, which can be derived rigorously from
the Fokker-Planck equation, only apply to tidally truncated
collisional stellar systems such as globular clusters).  It is seen on
Fig. 3 of Fa \& Pedron (2001) that an isothermal distribution in the
inner region provides a sensibly better fit to observations than a
polytropic one (it can reproduce the oscillations). In addition, the
model of Hjorth
\& Madsen (1993) has only one fitting parameter $\Phi_{0}$ (the central potential) instead of two $\Phi_{0}$ and $q$ in Fa \& Pedron's analysis. These remarks tend to show that generalized thermodynamics is not necessary to understand the structure of self-gravitating systems. A similar conclusion has been reached by Brands {\it et al.} (1999) in two-dimensional turbulence.}. On the other 
hand, the ``gravothermal catastrophe'' (Lynden-Bell \& Wood 1968)
should not throw doubts on the validity of statistical mechanical
arguments applied to self-gravitating systems. On the contrary, it is
gratifying that standard thermodynamics is consistent with the natural
tendency of self-gravitating systems to undergo a gravitational
collapse (Jeans instability).  The gravothermal catastrophe has been
confirmed by sophisticated numerical simulations (see, e.g., Larson
1970, Cohn 1980) based on standard kinetic theory
(Landau-Fokker-Planck equations) and it is expected to be at work in
globular clusters. The existence of negative specific heats in the
microcanonical ensemble (and the resulting inequivalence of
statistical ensembles) is a direct consequence of the Virial theorem
for self-gravitating systems and is not in contradiction with
ordinary thermodynamics (Lynden-Bell \& Lynden-Bell 1977, Padmanabhan
1990). Finally, self-gravitating systems have a well-defined (albeit
unusual) thermodynamical limit in which the number of particles $N$
and the volume $R^{3}$ go to infinity keeping $N/R$ fixed (de Vega \&
Sanchez 2001).

Therefore, we are tempted to beleive that ordinary statistical
mechanics is still relevant for describing self-gravitating systems
and two-dimensional vortices. An isothermal distribution corresponds
to the most probable distribution reached by a system after a complex
evolution during which microscopic information is lost. Accordingly,
it maximizes the Boltzmann entropy, which is a measure of the number
of microstates associated with a given macrostate (see, e.g.,
Ogorodnikov 1965, Lynden-Bell 1967). This definition of the entropy
does not depend whether the particles are in interaction or
not. Nonlocality intervenes only when the entropy is maximized for a
given value of energy. It is implicitly assumed, however, that all
accessible microstates are equiprobable, which is the fundamental
postulate of statistical mechanics.  Generalized thermodynamics is
interesting for extending standard results of thermodynamics to a
larger class of functionals which possess nice mathematical
properties, but it should not be presented (in our point of view) as
an improvement or a replacement of classical thermodynamics. We agree
with Tsallis and co-workers that the kinetic theory of
self-gravitating systems (see Kandrup 1981) and two-dimensional
vortices (see Chavanis 2001c) encounters some problems due to the
occurence of memory effects, spatial delocalization and logarithmic
divergences in the diffusion coefficient. For these reasons, and also
for the problems of incomplete relaxation (lack of ergodicity)
mentioned previously, the system may not necessarily reach the maximum
entropy state described by the Boltzmann distribution, even in the
mixing region.  The deviation from the Boltzmann distribution is never
too severe (see, e.g., the isothermal core of globular clusters and of
elliptical galaxies) so that ordinary statistical mechanics provides a
fairly good {\it prediction} from any initial condition. It may happen
that the effective equilibrium distribution can be {\it fitted} by the
$q$-distribution proposed by Tsallis and co-workers better than by the
Boltzmann distribution. This is the case for {\it some} examples of
statistical equilibrium in two-dimensional turbulence (see the
discussion of Brands {\it et al.}  1999). However, this agreement may
be fortuitous rather than dictated by a general physical principle
since there is a parameter $q$ in the theory which, in practice, must
be {\it adjusted} in each case. In fact, it is advocated by Tsallis
(private communication) that this parameter is not free but uniquely
determined by the microscopic dynamics of the system. It can be
relatively easy to determine if the system is simple enough but it can
also be very hard to determine in the case of complex systems like
those involved in fluid mechanics and stellar dynamics. In that case,
it must be regarded as a fitting parameter. Unfortunately, the power
of prediction of Tsallis theory is limited as long as a precise
prescription for determining $q$ is not given. Tsallis entropy could
be considered, however, as a {\it heuristic} attempt to take into
account non ergodic effects in complex systems. In view of these
remarks, it would be of considerable importance to see if the non
Markovian and non local generalized kinetic equations proposed by
Kandrup (1981) in stellar dynamics and by Chavanis (2001c) in vortex
dynamics allow for a stationary distribution of the form conjectured
by Tsallis, instead of the Boltzmann distribution. From our point of
view, we beleive that they do {\it not} select any ``universal''
distribution except in the approximation where (i) memory effects are
ignored (ii) a local approximation is made (in the stellar context)
(iii) ergodicity is assumed, in which case the classical Boltzmann
distribution is obtained. It is hard to beleive that extremely
complicated effects of non ergodicity, spatial delocalization and
memory can be encapsulated in a simple functional with a prescribed
value of $q$. However, if it can be proved that these generalized
kinetic equations rigorously converge towards a Tsallis distribution
with $q\neq 1$, this would of course be a strong argument in favour of
generalized thermodynamics. Unfortunately, the study of these
generalized kinetic equations seems of considerable difficulty and
will demand a long standing effort.

\section{Conclusion}
\label{sec_conclusion}

This study has revealed that the striking results obtained by Antonov (1962) and Lynden-Bell \& Wood (1968) in their investigations on the thermodynamics of self-gravitating systems are not restricted to isothermal configurations but also occur for polytropic configurations. It is possible that this formal analogy be the mark of a generalized thermodynamics (Tsallis entropy) but this point needs to be discussed in greater detail. In any case, the study presented in this paper and in our companion papers (Chavanis 2001a,b) provide a unified description of the structure and stability of isothermal and polytropic gas spheres. This is a useful complement to the existing literature on the subject (Chandrasekhar 1942). Quite remarkably, the stability analysis can be performed analytically or by using graphical constructions. Our study is limited in its applications by the requirement of a confining box in the case of isothermal configurations and polytropes of index $n\ge 5$. However, this box model (first introduced by Antonov) should be sufficient to exhibit the essential features of these systems and it is also very convenient for a theoretical analysis. In future works (in preparation), we shall extend our analytical results to the case of composite configurations made of an isothermal core and a polytropic envelope with index $n<5$. This is an important generalization for a better description of self-gravitating systems from the viewpoint of statistical mechanics and thermodynamics.

\section{Acknowledgments}
\label{sec_ack}

I acknowledge interesting comments from C. Tsallis and J. Perez on the first draft of this paper. This work was initiated during my stay at the Institute for Theoretical Physics, Santa Barbara, during the program on Hydrodynamical and Astrophysical Turbulence (February-June 2000). This research was supported in part by the National Science Foundation under Grant No. PHY94-07194.

\end{document}